\documentstyle[12pt]{article}
\begin{document}

\renewcommand{\baselinestretch}{1.5}
\newcommand\beq{\begin{equation}}
\newcommand\eeq{\end{equation}}

\centerline{\bf Integers and Fractions}
\vskip 1 true cm

The 1998 Nobel Prize in Physics has been awarded jointly to Professor Robert 
B. Laughlin, Stanford University, USA, Professor Horst L. St\"ormer, 
Columbia University and Bell Labs, USA, and Professor Daniel C. Tsui, 
Princeton University, USA, for "their discovery of a new form of quantum 
fluid with fractionally charged excitations" appearing in the fractional 
quantum Hall effect. The quantum Hall effect has two versions, the integer 
(which won an earlier Nobel Prize) and the fractional. These two versions 
share the remarkable feature that an experimentally measured quantity, 
characterising one aspect of a complicated many-particle system, stays
perfectly fixed at some simple and universal values even though many 
different parameters of the system vary from sample to sample (in fact,
some parameters like disorder vary quite randomly). In addition, the 
fractional quantum Hall effect has the astounding property that the low-energy 
excitations of the system carry a charge which is a simple fraction of the 
charge of an electron, even though the system is composed entirely of particles 
like atoms, ions and electrons, all of whose charges are integer multiples of 
the electronic charge. This article will briefly review the quantum Hall 
effect and the contributions of this year's Noble Laureates. A few references 
are given at the end for the reader who wants to learn more about the subject.

In 1879, Edwin Hall discovered that when a two-dimensional layer of electrons
(such as in a thin gold plate) is placed in a magnetic field perpendicular to 
the layer (pointing, say, in the $\hat z$-direction), an electric field along 
one direction in the layer (say, the $\hat y$-direction) causes a current flow
along the $\hat x$-direction which is at right angles to both
the electric and the magnetic field. The ratio of the current 
density $J_x$ to the electric field $E_y$ is now called the Hall conductivity
$\sigma_{xy}$. This quantity depends on the strength of the magnetic field,
and the density and sign (negative for electrons and positive for holes)
of the charge carriers. Classically, the Hall effect can be understood as 
being caused by the Lorentz force which acts on a charged particle moving in 
a magnetic field; this force acts in a direction perpendicular
to both the velocity and the magnetic field and is proportional to the
product of the two. The Hall conductivity is to be distinguished from the 
usual conductivity $\sigma_{xx}$ which denotes the ratio of the current 
density to the electric field along the {\it same} direction.

Hall performed his experiments at room temperatures and at magnetic fields
of less than a Tesla ($10^4$ Gauss). By the end of the 1970's, semiconductor 
technology had progressed to the extent that experimenters could perform 
similar measurements at the interface of a semiconductor and an insulator 
(such as a $Si - SiO_2$ device called a MOSFET) at very low temperatures of 
around $1^o K$ in magnetic fields of around $10$ Tesla. The two-dimensional 
interface acts as a quantum well which is very narrow (typically, $5 \times 
10^{-7}$ cm wide) in the direction perpendicular to the interface. This freezes 
out the motion of the electrons (or holes) in that direction thereby 
constraining them to move only along the two dimensions parallel to the
interface. The density of carriers in these systems is about $10^{11}$
cm$^{-2}$, while the mobility of the carriers is typically $10^4$ 
cm$^2$/Volt-sec. We recall here that the mobility denotes the drift velocity 
achieved by a carrier when an electric field of unit strength is 
applied. The finite value of the mobility in the above systems is due to
the various imperfections and impurities which cause the carriers to get 
scattered every $10^{-12}$ sec or so. 

In 1980, the German physicist Klaus von Klitzing and his
collaborators observed in such systems that the Hall conductivity 
$\sigma_{xy}$ does not vary linearly with the strength of the magnetic field 
(as the classical argument would suggest) but rather shows plateaus, i.e., 
$\sigma_{xy}$ does not vary at all for certain ranges of the applied
magnetic field. Further, its value at the plateaus is quantised at the
remarkably simple values
\beq
\sigma_{xy} ~=~ r ~\frac{e^2}{h} ~,
\eeq
where $r$ is a small integer like $1,2,3,...$ (whose value depends on the 
density of the carriers and the magnetic field), $e$ is the charge of an
electron, and $h$ is $2\pi$ times the Planck's constant $\hbar$. (The value 
of $\sigma_{xx}$ is extremely close to zero wherever $\sigma_{xy}$ has 
a plateau). Amazingly, the quantised 
values of $\sigma_{xy}$ do not depend at all on any other parameters of the
system such as its shape and size, the effective masses of the carriers, the 
strengths of their interactions with each other and with impurities, and the 
density of impurities. Finally, the plateau
values of $\sigma_{xy}$ have an experimental
uncertainty of only about one part in ten million which makes it one of the 
most accurately measured quantities in all of science. For this reason, the
integer quantum Hall effect (IQHE), so called because $r$ is an integer, is 
now used to define the value of $e^2 /h$ which is a fundamental constant of 
nature. The 1985 Nobel Prize in Physics was awarded to von Klitzing for his
discovery of the IQHE.

The IQHE can be explained by invoking some facts from one-particle quantum 
mechanics. A magnetic field leads to the formation of energy bands called 
Landau levels which are separated from each other by gaps. The number of states 
in each Landau level is directly proportional to both the area of the system 
and the strength of the magnetic field. Due to the
Pauli exclusion principle for fermions, each state can accommodate only one 
electron; at very low temperatures, all states up to a maximum energy called
the Fermi energy are occupied. If the density of electrons is such that 
exactly an integer number of Landau level is filled, then the 
Hall conductivity is given by that integer times $e^2/h$. Further, 
the existence of a finite energy gap to the next (unoccupied) 
Landau level would lead
to the vanishing of the usual conductivity $\sigma_{xx}$ at a plateau at zero
temperature; if the temperature is finite but much smaller than the gap, then
$\sigma_{xx}$ is exponentially small. 

The above simplified discussion has to be modified in a major way to take 
into account the imperfections and impurities which are randomly located in 
the system; we will refer to all these point objects as disorder. In an ideal
system with no disorder, 
each Landau level would be infinitely narrow and all the states would be 
spatially extended. In a real system, the disorder broadens each Landau level, 
and also converts a number of states from extended states (which can carry 
current) to localised states (which cannot carry current). We thus get 
alternating and non-overlapping bands of extended and localised states, with 
the extended bands staying close to the energies of the ideal Landau levels. 
The existence of localised bands intially led people to worry that the 
Hall conductivity may depend 
sensitively and unpredictably on the amount of disorder. However, using some 
clever arguments based on gauge invariance and topology, theorists were able 
to prove the amazing result that the extended band in any given Landau level 
contributes {\it exactly} the same amount to the Hall conductivity as that 
entire Landau level would have contributed if there had been no disorder and 
all the states had been extended. This means that if the Fermi energy lies 
within a band of localised states, and there is an integer number of 
filled bands of extended states below that energy, then the Hall 
conductivity is simply 
equal to that integer times $e^2 /h$. Thus, although the localised states do 
not contribute to the Hall conductivity, they are crucial for the finite widths
of the conductivity plateaus; this is because the Hall conductivity does not 
change at all if the Fermi energy moves by small amounts (due to changes in 
the carrier density or magnetic field), as long as it stays within a given 
localised band.

In 1982, St\"ormer, Tsui and collaborators working on some other systems
discovered additional plateaus in $\sigma_{xy}$ of the same form as in Eq. (1),
but with the parameter $r$ now equal to some simple odd-denominator fractions 
like $1/3, 2/3, 1/5, 2/5, 3/7, ...$; simultaneously, $\sigma_{xx}$ vanishes. 
Note that $r$ denotes the number of Landau levels occupied by electrons, hence
a value of $r$ less than $1$ means that the lowest Landau level is only
partially filled. Although this {\it fractional} quantum Hall effect 
(FQHE) sounds similar to the IQHE, both its discovery and its
theoretical understanding involved much greater levels of difficulty and
subtlety than the IQHE. Experimentally, the carrier mobility required to
observe the FQHE is many times times larger, i.e., about $10^5 - 10^6$ cm$^2$/
Volt-sec. This required the use of a new class of systems such as $GaAs - 
AlGaAs$ heterostructures. A technique called modulation doping
molecular beam epitaxy was used to produce interfaces which are extremely 
sharp, almost free of defects and well-separated from the impurities
(namely, the dopant atoms which act as donors or acceptors). This
vastly reduces the rate of scattering of the carriers which reside at the 
interface and therefore leads to much higher mobilities. The FQHE also 
requires much lower temperatures going down to about $0.1 ^o K$ and higher 
magnetic fields up to about $30$ Tesla (since the filling $r$ is inversely
proportional to the magnetic field).

The discovery of the FQHE raised the theoretical question of why there should
be a gap above the ground state (as is suggested by the vanishing of 
$\sigma_{xx}$) when the lowest Landau level is only partially occupied by
electrons. Clearly, this requires an explanation which is very different from
the one for the IQHE. Within an year, Laughlin came up with a many-body
explanation for the existence of a gap. In doing so, he also discovered a very
accurate variational quantum mechanical wave function for the system which 
explained all the observed phenomena and predicted some new ones. Laughlin's 
key idea was that the Coulomb repulsion between the carriers (assumed to be 
electrons from here on) must play a crucial role in the
FQHE unlike the case of the IQHE. This must lead to the electrons "avoiding"
each other as best as they can, subject to the constraint that they are all
in the lowest Landau level. Using some general properties of states in the
lowest Landau level, Laughlin introduced a variational many-electron wave 
function of the form
\beq
\Psi (z_j , z_j^\star ) ~=~ \prod_{j<k} ~( z_j ~-~ z_k )^m ~\exp ~[- \sum_j
z_j z_j^\star ~] ~,
\eeq
where $z_j = x_j + i y_j$ and $z_j^\star = x_i - i y_j$ are complex variables
denoting the position of electron $j$ in the $x-y$ plane; to simplify the
notation, we have scaled the coordinates in some
way so as to make them dimensionless. For $m$ equal to an odd integer, this
wave function is antisymmetric (as the wave function of several identical
fermions should be; the spin degree of freedom plays no role here since the
electrons are completely polarised in the lowest Landau level), vanishes 
very rapidly as any two particles approach each other (thereby ensuring that 
they stay away from each other and minimise their Coulomb repulsion), and,
very importantly, describes a state with an uniform density 
corresponding to the special value of Landau level filling given by $r = 1/m$. 
(For other filling fractions which are not of the form $1/m$, Laughlin and
others suggested related variational wave functions using similar arguments). 
The state defined by Eq. (2) describes a peculiar kind of quantum liquid which 
has no long-range positional order (unlike the atoms in a crystalline solid), 
but does have a long-range phase correlation; this is related to the fact that 
the wave function picks up a phase of $2 \pi m$ when any particle $j$ is 
taken in a loop around particle $k$, no matter how large the loop is.

After showing that the wave function in (2) perfectly explains the properties 
of the ground state such as the conductivity quantisation, Laughlin went on to 
study the nature of the low-energy excitations called quasiparticles. He 
showed that these are separated from the ground state by an energy gap of 
about $1 ^o K$, and that the quasiparticle size is typically about $5 \times 
10^{-7}$ cm which is substantially smaller than the average distance between 
two electrons (about $3 \times 10^{-6}$ cm). There are two types of 
quasiparticles, called quasielectrons and quasiholes; in the neighbourhood of 
these objects, the local electron density is slightly larger or smaller, 
respectively, than the special value of density (or filling) given by the 
Laughlin wave function. Since each quasiparticle costs a finite energy, the 
system tends to oppose any attempts to change its electron density away from 
one of the special fillings of $1/m$; this property is called 
incompressibility and it is one of the striking features of the fractional 
quantum Hall system, sometimes called a Laughlin liquid. If a small
change in filling is forced upon the system by varying either the number of
electrons or the magnetic field, the system reacts by producing a small
number of quasielectrons or quasiholes which then go and bind to one of the
points of disorder; this allows the rest of the system to remain at the 
special filling, the Hall conductivity thereby remaining exactly at the 
quantised value. Thus the finite widths of the Hall plateaus again depend 
on the presence of a small amount of disorder. 

Laughlin also made the extraordinary prediction that the
quasiparticles should carry a charge which is a fraction of the
electron charge, e.g., the quasielectron and quasihole charges should be 
equal to $\pm e/m$, respectively, if $r = 1/m$. This prediction has been
directly verified in recent experiments which study tiny fluctuations (called
shot noise) in the current flowing across a narrow region of a fractional 
quantum Hall system. The amount of shot noise is directly proportional 
to the charge of the carriers, and the experiments conclusively proved that 
these are fractional. (Although the quasiparticles cannot move through the
bulk of a quantum Hall system because they get pinned by impurities, they do
move freely along the edges of such a system, and they can jump from one edge
to a nearby edge as shown by the shot noise experiments).

For those who are uneasy with the idea of fractional charge, we should 
emphasise that the quasiparticles in a fractional quantum Hall system are not 
elementary particles (like electrons or photons), but are collective 
excitations involving a large number of electrons. It was already known in 
another many-body system (polyacetylene chains) that composite objects may 
carry fractional charge. The FQHE has provided us with a new and wonderful 
example of objects with a fractional quantum number. Incidentally, it has 
been suggested by some theorists that these quasiparticles are neither 
fermions (obeying Fermi-Dirac statistics) nor bosons (obeying Bose-Einstein 
statistics), but that they obey a novel kind of quantum statistics (called 
fractional or anyonic statistics) which lies in between the two possibilities; 
however this has not yet been proved experimentally.

Laughlin's wave function has led to new ways of thinking about strongly
interacting many-body systems. For instance, if $r = 1/m$ where $m$ is odd, it 
is natural to think of a quantum Hall system as being made up of "composite 
bosons" which are combinations of electrons and $m$ elementary quanta 
of magnetic flux. These "bosons"
effectively see a zero magnetic field, since the entire magnetic field has
been "absorbed" in the way these strange particles are defined. Now one can
view the FQHE as resulting from a Bose-Einstein condensation of these 
bosons at low temperatures, in the same way that Cooper pairs of electrons 
condense in superconductors or the atoms of $He^4$ condense to form a 
superfluid at low temperatures. This idea of composites of electric 
charge and magnetic flux has turned out to be remarkably useful in explaining 
many other aspects of quantum Hall systems, such as the behaviour of the 
system at the filling $r=1/2$ (which does {\it not} show a conductivity
plateau, but has some other interesting properties which are well-explained by
the notion of "composite fermions").

Before ending, one should appreciate that the theoretical and experimental 
work on the quantum Hall effect is a reflection of the high 
level of both industrial and university research and the close collaboration 
between the two in the West, particularly in the USA. 

\vskip .5 true cm

\vskip .5 true cm

\noindent ACKNOWLEDGEMENTS. I thank B. Sriram Shastry for some stimulating
discussions.

\vskip 1 true cm
\rightline{Diptiman Sen}
\rightline{\it Centre for Theoretical Studies,}
\rightline{\it Indian Institute of Science,}
\rightline{\it Bangalore 560012, India}


\begin{thebibliography}{99}
\vskip .5 true cm

\bibitem{a} Tsui, D. C., Stormer, H. L. and Gossard, A. C., {\it Physical 
Review Letters}, 1982, {\bf 48}, 1559-1562.

\bibitem{b} Laughlin, R. B., {\it Physical Review Letters}, 1983, {\bf 50}, 
1395-1398.

\bibitem{c} Halperin, B. I., The quantized Hall effect, {\it Scientific 
American}, April 1986, {\bf 254}, 40-48.

\bibitem{d} Eisenstein, J. P. and Stormer, H. L., The fractional quantum Hall 
effect, {\it Science}, 22 June 1990, {\bf 248}, 1510-1516. 

\bibitem{e} Kivelson, S., Lee, D.-H. and Zhang, S.-C., Electrons in flatland, 
{\it Scientific American}, March 1996, {\bf 274}, 64-69.

\bibitem{f} Anderson, P. W., When the electron falls apart, {\it Physics 
Today}, October 1997, {\bf 50}, 42-47.

\bibitem{g} Collins, G. P., Fractionally charged quasiparticles signal their 
presence with noise, {\it Physics Today}, November 1997, {\bf 50}, 17-19.

\bibitem{h} Daviss, B., Splitting the electron, {\it New Scientist}, 31 
January 1998, {\bf 157}, 36-40.

\end{thebibliography}
\end{document}